\documentclass[galaxies,article,accept,pdftex,moreauthors]{Definitions/mdpi}

\usepackage[labelformat=simple]{subcaption}

\DeclareCaptionLabelFormat{subcaptionlabel}{\normalfont(\textbf{#2}\normalfont)}
\newcommand{\Vmic}{$V_{\rm mic}$}

\newcommand{\teff}{$T_{\rm eff}$}
\newcommand{\kms}{km\,s$^{-1}$}
\newcommand{\lgg}{$\log{g}$}

\newcommand{\vs}{$v_{\rm e}\sin i$}
\newcommand{\abun}{$\log\varepsilon$}
\def\lgt{$\log\tau_{5000}$}
\def\gL{$g_{\rm L}$}
\newcommand{\SV}{\textsc{SynthVb}}
\newcommand{\SM}{\textsc{Synmast}}

\newcommand{\vald}{\textsc{VALD}}

\newcommand{\DETAIL}{\textsc{DETAIL}}
\def\ione{\,{\sc i}}
\def\ii{\,{\sc ii}}
\def\iii{\,{\sc iii}}
\def\bz{$\langle B_{\rm z} \rangle$}
\def\br{$\langle B_{\rm r} \rangle$}
\def\bt{$\langle B_{\rm t} \rangle$}
\PassOptionsToPackage{unicode}{hyperref}
\PassOptionsToPackage{naturalnames}{hyperref}

\firstpage{1} 
\pubvolume{1}
\issuenum{1}
\articlenumber{0}
\pubyear{2024}
\copyrightyear{2024}
\externaleditor{Academic Editor: Dominic Bowman}
\datereceived{23 August 2024} 
\daterevised{19 September 2024} 
\dateaccepted{22 September 2024} 
\datepublished{ } 
\hreflink{https://doi.org/} 

\Title{Slowly Rotating Peculiar Star BD+00$^\circ$1659 as a Benchmark for Stratification Studies in Ap/Bp Stars}

\TitleCitation{Slowly Rotating Peculiar Star BD+00$^\circ$1659 as a Benchmark for Stratification Studies in Ap/Bp Stars}

\Author{Anna Romanovskaya *\orcidA{}, 
Tatiana Ryabchikova \orcidB{}, Yury Pakhomov
\orcidC{}, Ilya Potravnov \orcidD{} and Tatyana Sitnova \orcidE{}}

\AuthorNames{Anna Romanovskaya, Tatiana Ryabchikova, Yury Pakhomov, Ilya Potravnov and Tatyana Sitnova}
\AuthorCitation{Romanovskaya,
A.; Ryabchikova, T.; Pakhomov, Y.; Potravnov, I.; Sitnova, T.}

\address[1]{%

Institute of Astronomy of Russain Academy of Sciences, Moscow 119017, Russia, 
 Current Address: Pyatnitskaya Str. 48, Moscow 119017, Russia.; 
ryabchik@inasan.ru (T.R.); pakhomov@inasan.ru (Y.P.); ilya.astro@gmail.com (I.P.); sitamih@gmail.com (T.S.)}

\corres{\hangafter=1 \hangindent=1.05em \hspace{-0.82em}Correspondence: annarom@inasan.ru}

\abstract{We present the results of a self-consistent analysis of the magnetic silicon star BD+00$^\circ$1659, based on its high-resolution spectra taken from the ESPaDOnS archive ($R$ = 68,000). This narrow-lined star shows the typical high Si abundance and Si\ii--\iii\ anomaly, making it an ideal prototype for investigating the vertical distribution of Si and Fe in the stellar atmosphere. The derived abundances, ranging from helium to lanthanides, confirm the star's classification as a silicon Bp spectral type. Silicon and iron are represented by lines of different ionisation stages (Fe\ione--\iii, Si\ione--\iii), indicating an ionisation imbalance interpreted as evidence of atmospheric stratification. Our stratification analysis reveals that there is a jump in iron and silicon abundances of 1.5 dex at atmospheric layers with an optical depth of \lgt\ = $-$0.85--$-$1.00. Non-LTE calculations for iron in this stratified atmosphere show minor non-LTE effects. Our results can be applied to studying the impact of stratification on the emergent flux in rapidly rotating Si stars with similar atmospheric parameters and abundance anomalies (for example, MX TrA), where direct stratification analysis is challenging due to line~blending.}


\begin{document}


\section{Introduction}\label{Intro}

Magnetic chemically peculiar (Ap) stars are characterised by the presence of global magnetic fields and atmospheric abundance anomalies, particularly an excess of Si, Cr, and rare-earth elements (REE) ranging from 1 to 4 dex above solar values. It is believed that the magnetic field suppresses mixing processes in the atmospheres of magnetic stars, thereby creating conditions favourable for element separation. Michaud \cite{1970ApJ...160..641M} proposed that the observed abundance anomalies could be explained by the diffusion of atoms and ions due to the combined effects of gravitational settling and radiation levitation. When gravitational pressure is the dominant force, elements tend to settle into deeper layers of the atmosphere. When, in contrast, radiation pressure dominates, elements concentrate in the upper atmosphere. This element separation results in vertical abundance gradients, known as stratification, which are responsible for the observed abundance anomalies. Stratification alters the atmospheric structure through changes in line and continuum opacities, leading to differences in the observed energy distribution compared to chemically normal stars \cite{2009A&A...499..851K}. 

The observational evidence for stratification are: an impossibility to fit the wings and cores of strong lines with a single value of abundance; the dependence of abundances on the excitation potential of transition; and the abundance difference yielded by lines of different ions of the same element \cite{2003IAUS..210..301R}. The latter effect was observed for silicon lines in the spectra of Bp and even normal stars and has been referred to as the ``Si\ii--\iii\ anomaly''\,\cite{2013A&A...551A..30B}. While in normal stars the difference in Si\ii--\iii\ abundances can be almost completely reduced by accounting for non-LTE effects \cite{2020MNRAS.493.6095M}, in Bp stars, it is usually larger and has been suggested to be a result of stratification effects. Although Ap/Bp stars statistically rotate slower than normal stars of the same spectral types, the number of suitable objects is severely limited even among this subgroup. Significant line blending which complicates studies of stratification appears for velocities above \vs\,  $\approx$ 20 \kms. Hence, it is of practical interest whether the stratification profile derived in slowly rotating objects can be applied for the analysis of \mbox{faster ones.}

BD+00$^\circ$1659 is a magnetic chemically peculiar star located in the open cluster NGC 2301, in the Galactic plane \cite{1962ZA.....54...41G}. It is classified as a B9 Si star in the catalogue \cite{1991A&AS...89..429R}. Bailey and Landstreet \cite{2013A&A...551A..30B} investigated BD+00$^\circ$1659 among 16 more magnetic Bp stars with effective temperatures ranging from 11,000 to 15,000 K and different rotation velocities. These stars exhibit a silicon overabundance of up to 1--1.5~dex compared to the solar value and a discrepancy between the abundances derived from Si\ii\ and Si\iii\ lines, with the latter being approximately 1 dex higher. The dipole magnetic field of BD+00$^\circ$1659 was estimated to be 1200 G \cite{2013A&A...551A..30B}, although four \bz\ measurements over the period 2005--2008~\cite{2015AstBu..70..444R} suggest that the magnetic field is weak. Its mean value of $\sim$300~G  does not exceed significantly the measurement error of 100--130~G. Except for the pronounced ``Si\ii\--\iii\ anomaly''\, an important feature of the star is its low projection equatorial velocity \mbox{\vs\,  = 7 $\pm$ 1 \kms}~ \cite{2013A&A...551A..30B}. Therefore, we performed abundance and stratification analysis of BD+00$^\circ$1659 and consider whether it can be used as a benchmark for stratification studies in other Ap/Bp stars in the close temperature range.

The paper is organised as follows. Section~\ref{obs} provides information about observational data and codes used in our study. The abundance and stratification analysis is described in Sections~\ref{abund} and \ref{strat}. Comparison of our results with those from the literature is given in Section~\ref{disscussion}. Conclusions are presented in Section~\ref{conclusions}.

\section{Observations and Stellar Atmosphere Parameters}
\label{obs}
For our analysis, we used the high-resolution spectrum extracted from the ESPaDOnS spectrograph (Canada--France--Hawaii Telescope) archive\footnote{https://www.cadc-ccda.hia-iha.nrc-cnrc.gc.ca/en/cfht/ (accessed on 23 September 2024)} covering the spectral range \mbox{3700--10,500 \AA\,} with resolving power $R$ = 68,000 and signal-to-noise ratio $SNR = 182$ near 5500 \AA\ (proposal ID---08AC09, UT Date---2008-03-28, PI---J. Landstreet). The spectrum was automatically processed and the wavelength calibrated with the Libre-ESpRIT package~\cite{Donati_1997}.

\begin{figure}[H]

\begin{adjustwidth}{-\extralength}{0cm}
\centering 
\captionsetup[subfigure]{justification=centering}
 \begin{subfigure}[t]{1.0\textwidth}
\centering
     \includegraphics[width=\textwidth]{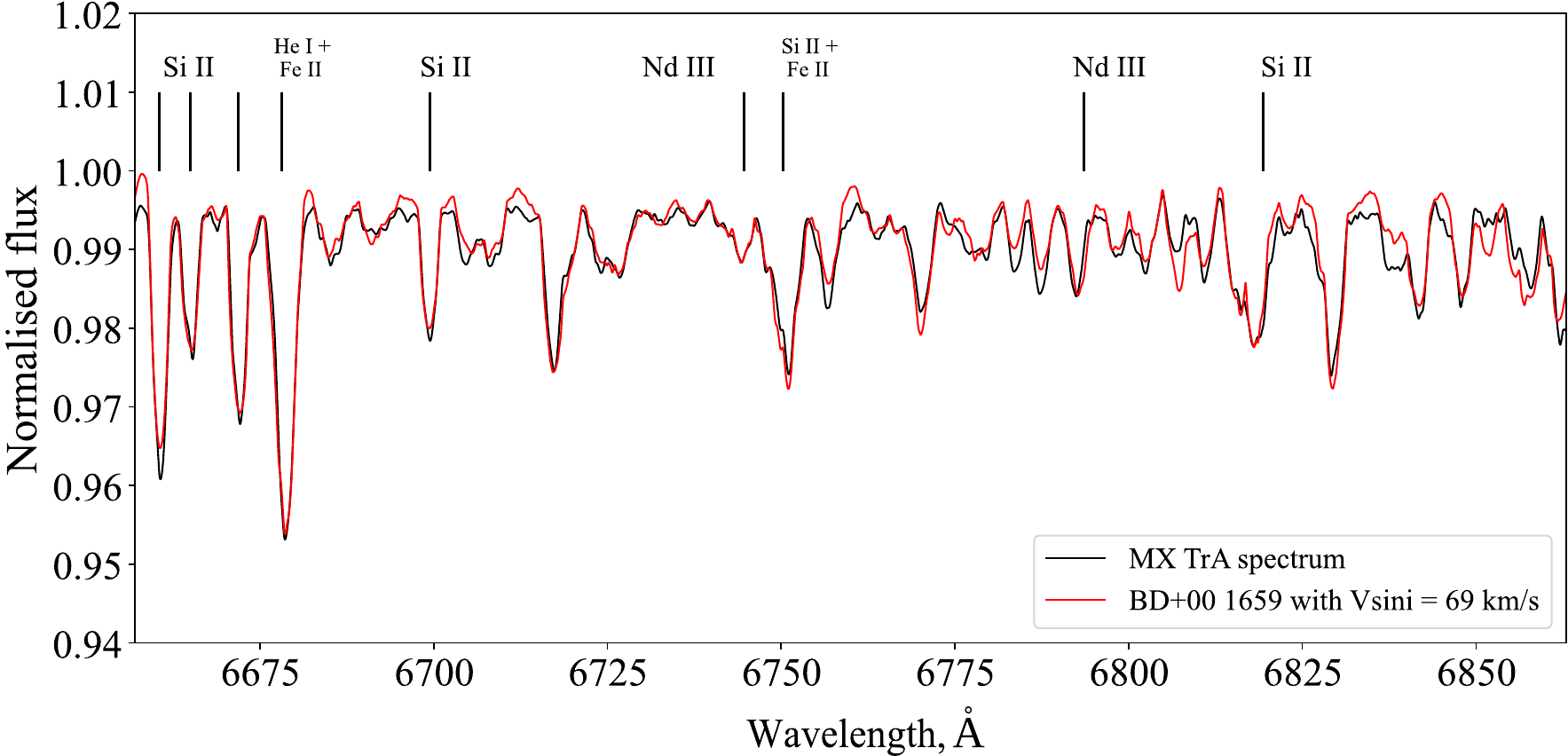}
     \caption{}
 \end{subfigure}

 \medskip
 \begin{subfigure}[t]{0.35\textwidth}
\centering
     \includegraphics[width=1.2\textwidth]{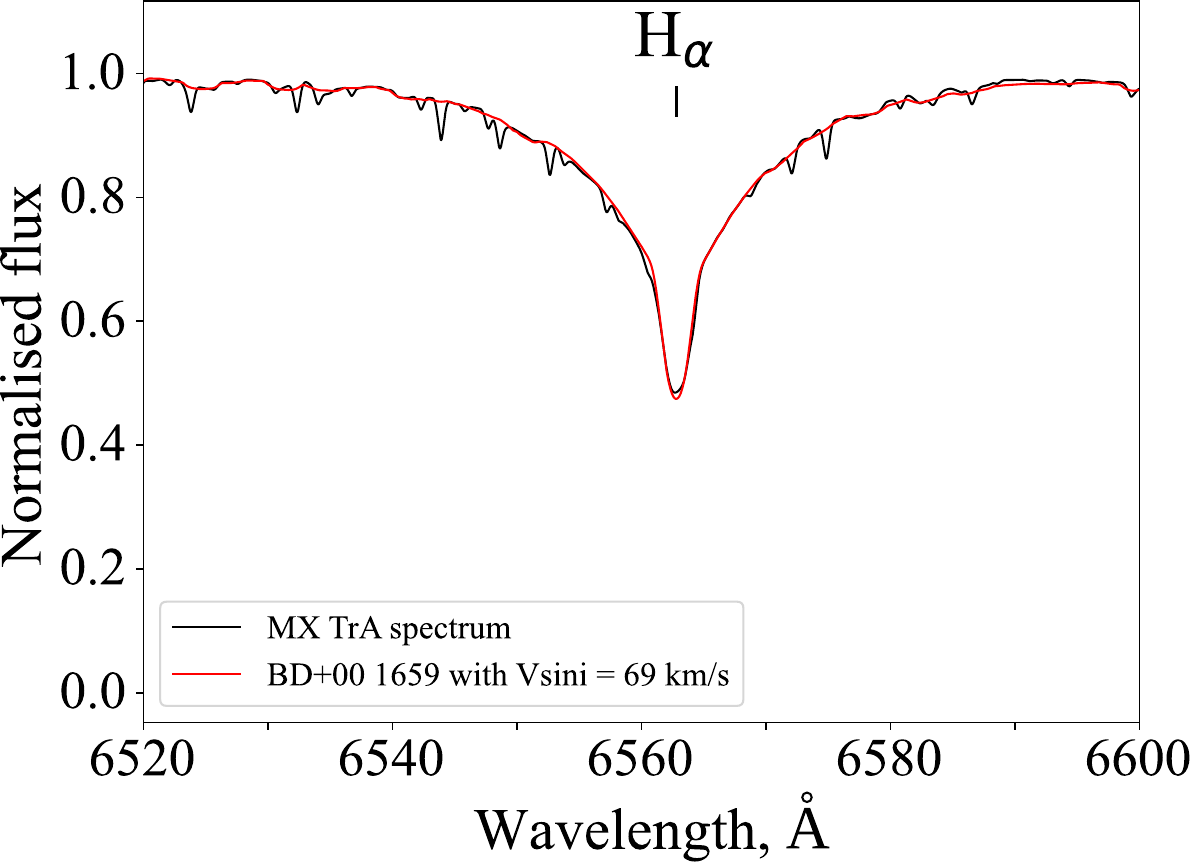}
 \end{subfigure}
  \quad\quad\quad
 \begin{subfigure}[t]{0.34\textwidth}
\centering
     \includegraphics[width=1.2\textwidth]{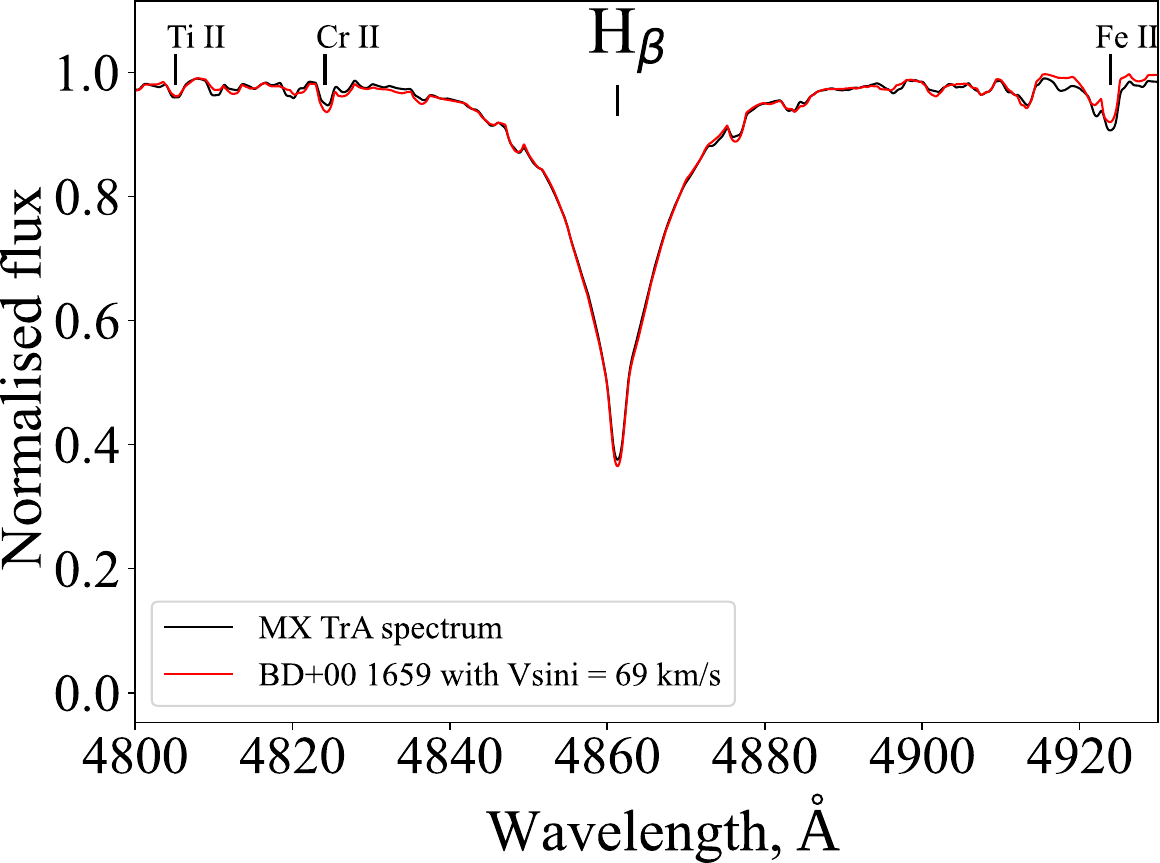}
     \caption{}
 \end{subfigure}
 \quad\quad\quad
 \begin{subfigure}[t]{0.34\textwidth}
\centering
     \includegraphics[width=1.2\textwidth]{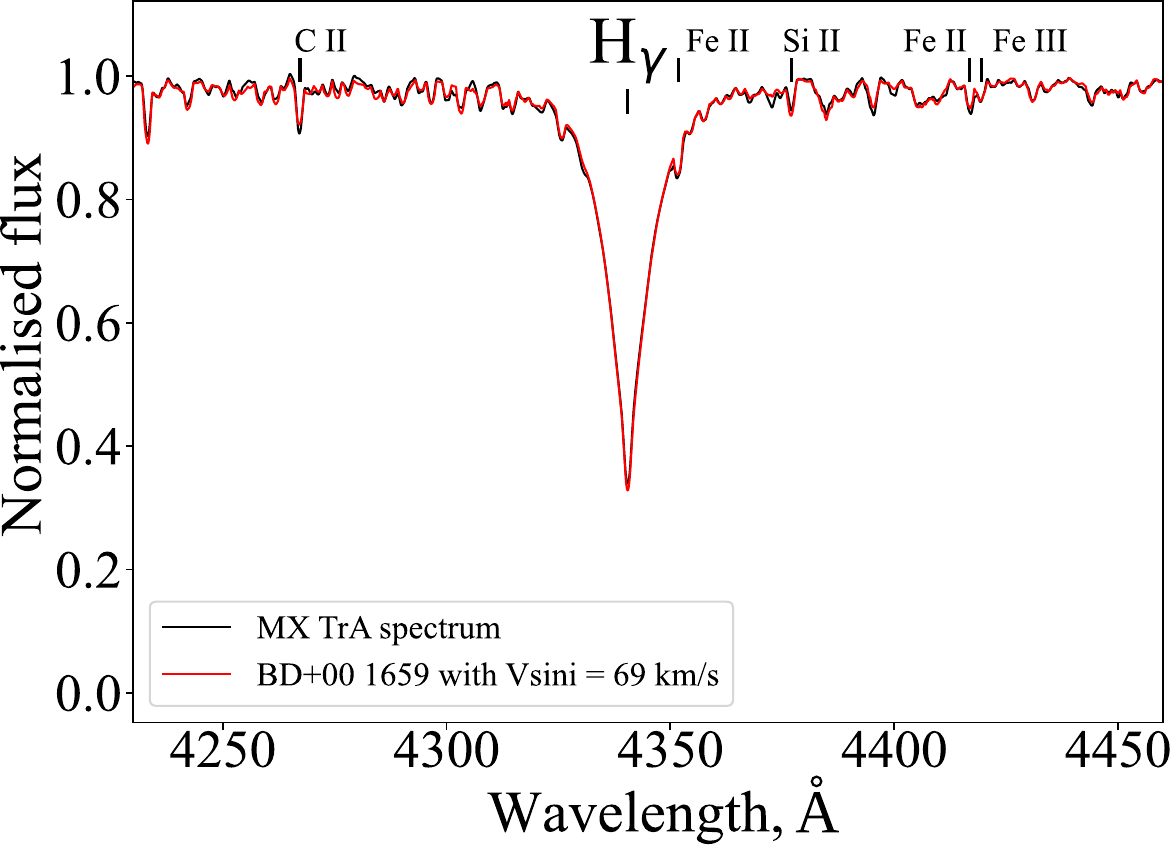}
 \end{subfigure}

    \medskip
\begin{subfigure}[t]{0.34\textwidth}
\centering
     \includegraphics[width=1.2\textwidth]{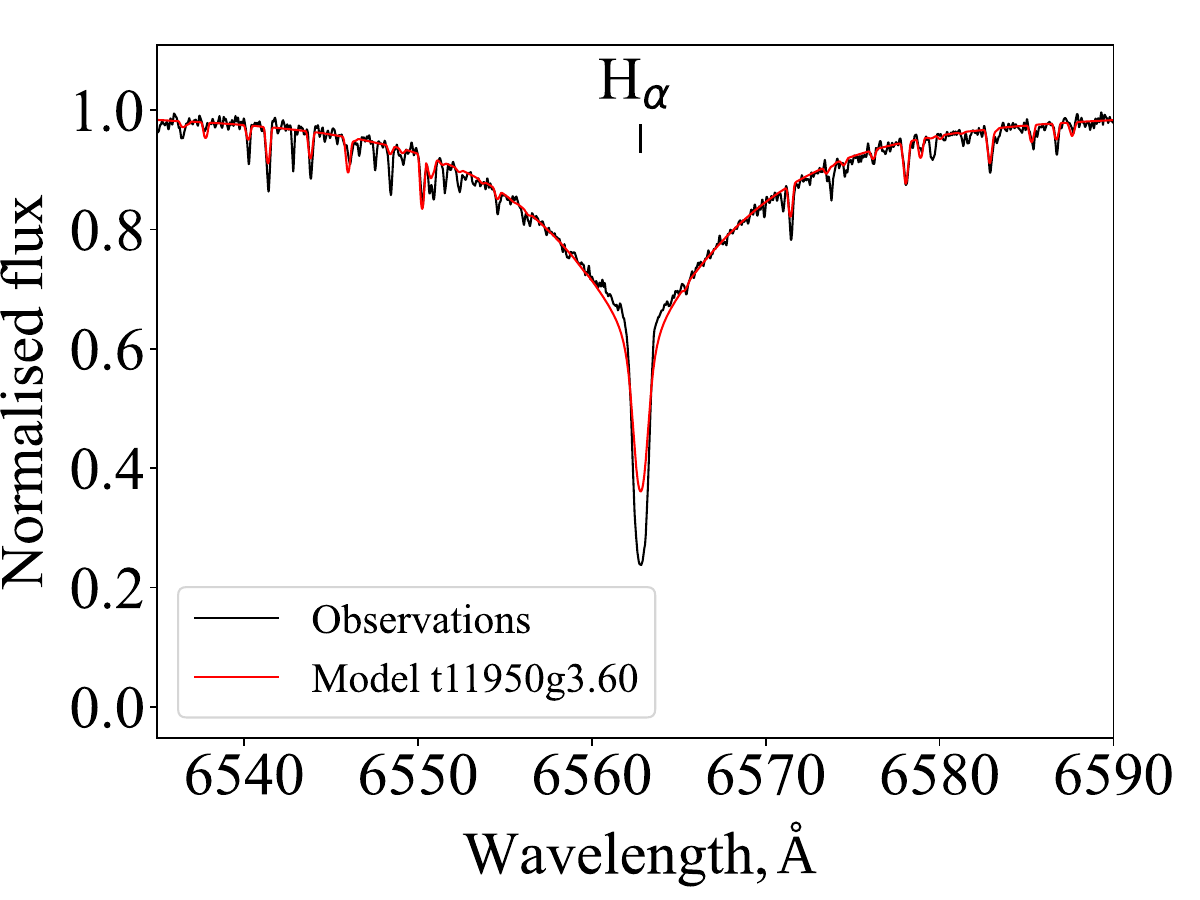}
 \end{subfigure}
 \quad\quad\quad
 \begin{subfigure}[t]{0.34\textwidth}
\centering
     \includegraphics[width=1.2\textwidth]{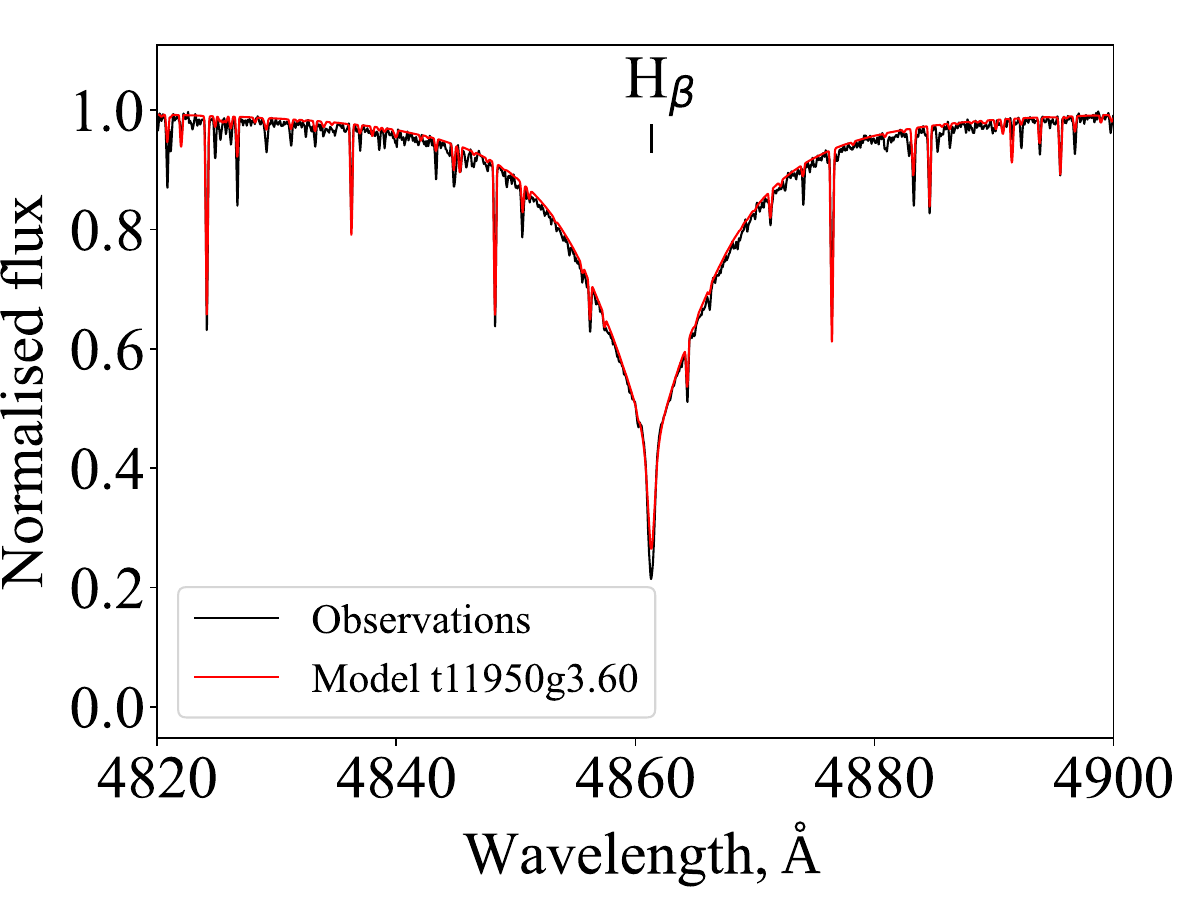}
     \caption{}
 \end{subfigure}
 \quad\quad\quad
 \begin{subfigure}[t]{0.34\textwidth}
\centering
     \includegraphics[width=1.2\textwidth]{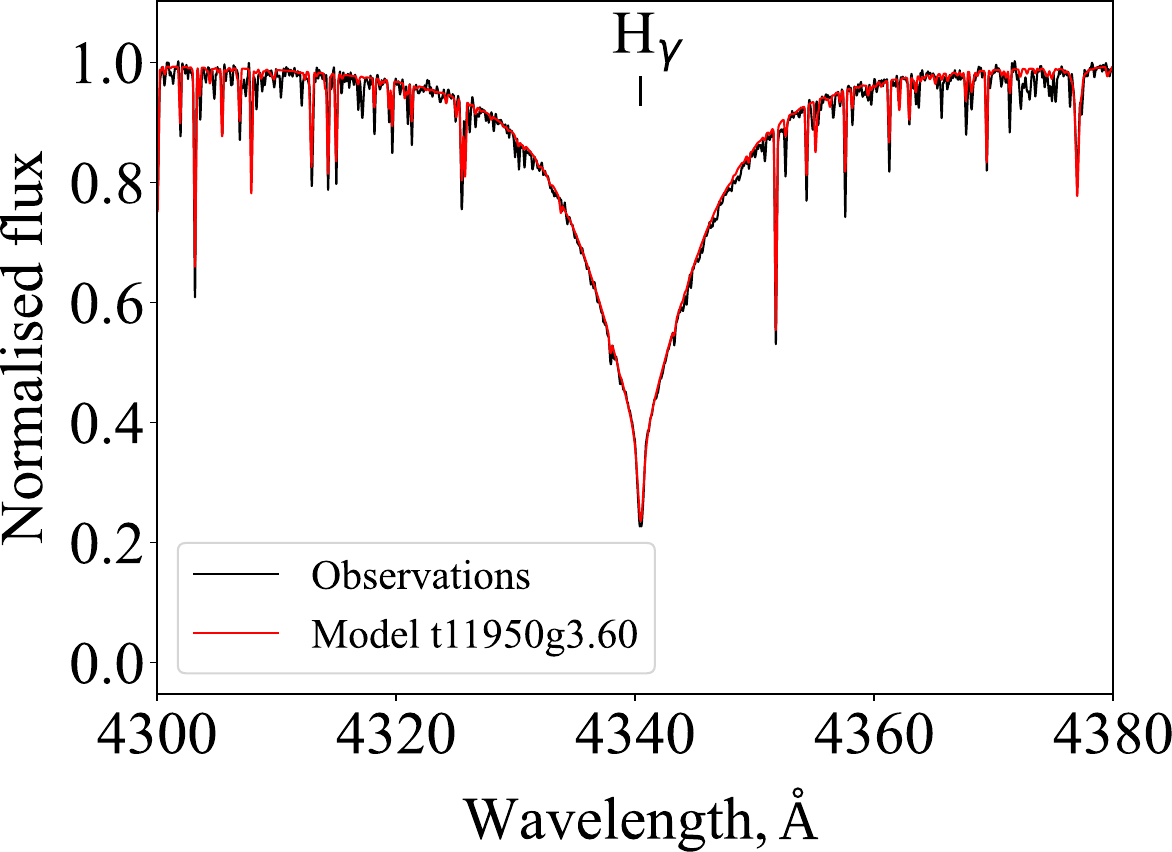}
 \end{subfigure}
\end{adjustwidth}
 
 \caption{Comparison between MX TrA (black line) and BD+00$^\circ$1659 (red line) spectra (\textbf{a},\textbf{b}). The observed spectrum of BD+00$^\circ$1659 is broadened with a rotational velocity of \vs\ = 69 \kms. (\textbf{a}) Spectral range 6670--6890 \AA\, with several Si and Fe lines, (\textbf{b}) hydrogen line profiles, (\textbf{c}) hydrogen line profiles in the observed spectrum of BD+00$^\circ$1659 (black line) and synthesis spectrum calculated with t11950g3.60 model (red line).}\label{fig1}
 \end{figure}

\begin{figure}[H]
\includegraphics[width=12.5 cm]{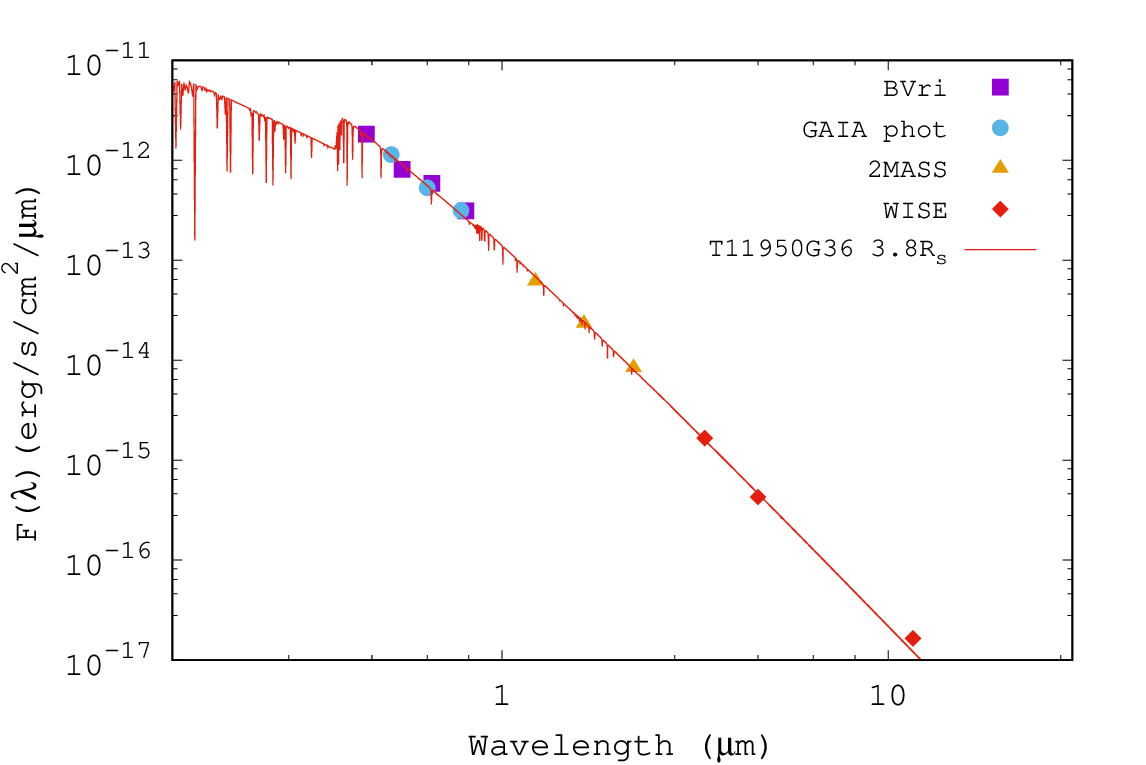}
\caption{Spectral energy distribution of BD+00$^\circ$1659 (points) compared with theoretical flux (solid line) from adopted model atmosphere.}
\label{SED}
\end{figure}

We performed atmospheric parameter determination using a self-consistent procedure based on spectral synthesis of Balmer lines profiles and spectral energy distribution (SED). Surprisingly, we found that BD+00$^\circ$1659 appears to be a twin of a previously studied ApSi star MX TrA (HD~152564) \cite{2024MNRAS.52710376P}. The observed spectra of BD+00$^\circ$1659 and MX TrA are nearly identical, particularly in the wavelength region around the hydrogen lines, when the observed spectrum of BD+00$^\circ$1659 is broadened with the rotational velocity \vs\ = 69 \kms\ of MX TrA (see Figure~\ref{fig1}a--c). The parameters of their atmospheres are: effective temperature \teff\ =11,950~K, surface gravity \lgg\ = 3.6, and microturbulence \Vmic\ = 0 \kms\ (likewise in other magnetic Ap/Bp stars). Theoretical SED, calculated with these parameters and scaled for the radius of MX TrA from \cite{2024MNRAS.52710376P}, align perfectly with the observational SED of BD+00$^\circ$1659 (Figure~\ref{SED}). Photometric observations of BD+00$^\circ$1659, collected via the Vizier-SED service\footnote{http://vizier.cds.unistra.fr/vizier/sed/ (accessed on 23 September 2024)}, were corrected for interstellar reddening with $E(B-V) = 0.03$~\cite{2005A&A...438.1163K} using Fitzpatrick's extinction curve \cite{2019ApJ...886..108F}. The distance to the star, $d = 864.38$ pc, was adopted from \cite{2021A&A...649A...1G}. Therefore, we used the pre-calculated model atmosphere of MX TrA from \cite{2024MNRAS.52710376P} (\teff\, = 11,950 K and \lgg\, = 3.60. Hereafter:  t11950g3.60) for the further analysis of BD+00$^\circ$1659.

The surface magnetic field of the star was analysed using the Fe\ii\ 6147.73 (Land\'e g-factor \gL = 0.83) and 6149.25 \AA\ (\gL = 1.35) lines, which are commonly used for magnetic field measurements. One more Fe\ii\ line at 6432.68 \AA\ (\gL = 1.83) was added. We computed the synthetic spectrum using the \SM\ code \cite{2007pms..conf..109K} with zero radial \br\ and transversial \bt\ components of the magnetic field vector and also allowing these components to vary. In the second case, we obtained a practically zero radial component and \bt~= 273 $\pm$ 198~G. Comparison of the spectral synthesis with single epoch observations used in our analysis revealed no signatures of magnetic line intensification exceeding the determination error. Therefore, subsequent synthetic spectrum calculations were performed using the nonmagnetic \SV\ code \cite{2019ASPC..518..247T}. \SV\ enables calculating the synthetic spectrum either in the local thermodynamic equilibrium (LTE) or by accounting for departures from the LTE (non-LTE approach) via pre-computed departure coefficients (the ratio between non-LTE and LTE atomic level populations).
 
\section{Abundance Analysis}
\label{abund}
The abundance for each spectral line was determined by fitting the synthetic line profile to the observed one using the BinMag visualization tool \cite{2018ascl.soft05015K}. Atomic data were taken from \vald3\ \cite{1995A&AS..112..525P, 2019ARep...63.1010P}. The non-LTE calculations are performed with a modified version of the \DETAIL\, code \citep{1981PhDT.......111G,detail} which includes the opacity package updated by \citet{2011JPhCS.328a2015P}. For this study, we updated the \DETAIL\, code to enable accounting for chemical element stratification.
 
We determined abundances of 30 elements from He to Er. For Fe and Si, the abundances were obtained from the lines of the neutral atoms, and first and second ions. For other elements up to Ba, the abundances were obtained from the lines of the neutral atoms and first ions, and for the rare-earth elements, we used the lines of the first and second ions. Element abundances are given in a standard designation, \abun\ = $\log(N_{\rm el}/N_{\rm tot})$ + 12.04, where N$_{\rm el}$ and N$_{\rm tot}$ are number densities of a given chemical element and total number of atoms, respectively. The full linelist with the line atomic parameters and individual line abundances is available online (see {\bf Data Availability Statement} for details).

\begin{figure}[H]
\includegraphics[width=12 cm]{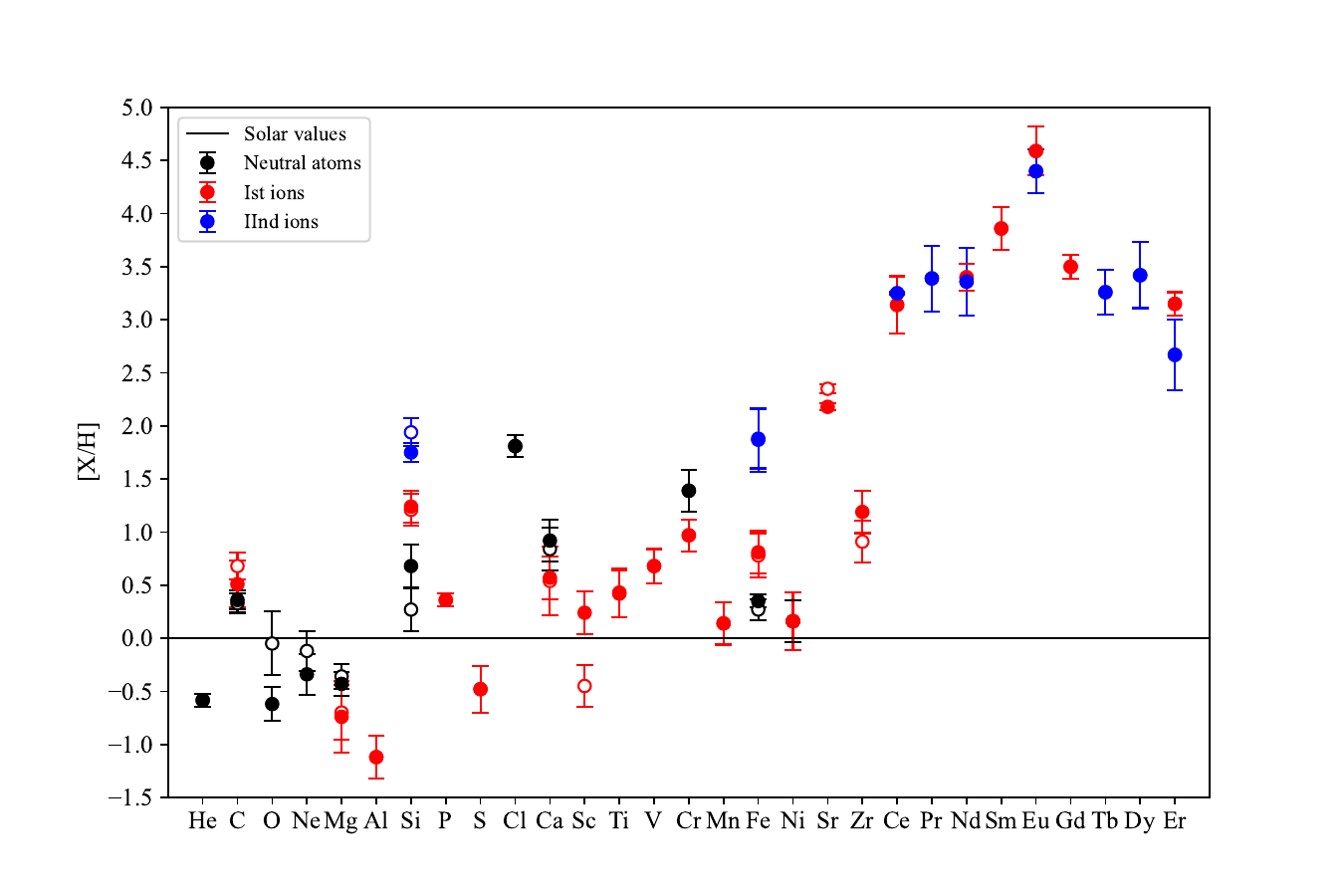}
\caption{Abundances of BD+00$^\circ$1659 relative to solar values. For those elements where non-LTE and LTE abundances are available, the corresponding LTE values are shown with open symbols.}
\label{fig3}
\end{figure}

He, Al, V, P, S, Cr, Mn, Ni, and REE abundances are derived in LTE. For C, O, Ne, Na, Mg, Si, Ca, Sc, Ti, Sr, Zr, and Ba, we determine LTE and non-LTE abundances using model atoms described in \cite{2016MNRAS.462.1123A, 2013AstL...39..126S, 2020ApJ...896...59A, 2014AstL...40..406A, 2018ApJ...866..153A, 2020MNRAS.493.6095M, 2018MNRAS.477.3343S, 2024MNRAS.527.8234M, 2020AstL...46..120S, 2020MNRAS.499.3706M}. Non-LTE calculations for Fe rely on a comprehensive Fe\ione-\ii-\iii\ model atom first introduced in \citet{2018IAUS..334..360S}. The derived abundances of BD+00$^\circ$1659 are presented in Table~\ref{MeanAbunds} and shown in Figure~\ref{fig3}. 

\begin{table}[H]
        \caption{Non-LTE and LTE abundances for atoms and ions in the atmosphere of BD+00$^\circ$1659.} 
\tablesize{\small}
\newcolumntype{R}{>{\raggedleft\arraybackslash}X}
        \begin{tabularx}{\textwidth}{L C |CCR|C|C}
\noalign{\hrule height 1.0pt}
                 \textbf{Ion} & & \multicolumn{3}{c|}{\textbf{BD+00$^\circ$1659}} & \textbf{MX TrA}  & \textbf{Sun}\\ 
                \hline     &  & \boldmath{\abun} & \textbf{[X/H]} & \textbf{n}\boldmath{$_{\rm l}$}  &\boldmath{\abun}&  \boldmath{\abun} \\
                \hline 
                He\ione    & L &10.34(06) & $-$0.584&  4 &10.43(09)   &10.924\\
                C\ione     & L & 8.80(09) &  0.33 &  5 &            & 8.47\\
                           & N & 8.83(09) &  0.36 &  5 &               &     \\
                C\ii       & L & 9.15(13) &  0.68 &  4 & 9.34(04)&     \\
                           & N & 8.98(22) &  0.51 &  4 &               &     \\
                O\ione     & L & 8.68(30) & $-$0.05 &  7 &           & 8.73\\
                           & N & 8.11(16) & $-$0.62 &  7 & 8.01(14) &     \\
\noalign{\hrule height 1.0pt}
\end{tabularx}

\end{table}
\begin{table}[H]\ContinuedFloat

\caption{\textit{Cont.}}
\newcolumntype{R}{>{\raggedleft\arraybackslash}X}
        \begin{tabularx}{\textwidth}{L C |CCR|C|C}
\noalign{\hrule height 1.0pt}
                 \textbf{Ion} & & \multicolumn{3}{c|}{\textbf{BD+00$^\circ$1659}} & \textbf{MX TrA}  & \textbf{Sun}\\ 
                \hline     &  & \boldmath{\abun} & \textbf{[X/H]} & \textbf{n}\boldmath{$_{\rm l}$}  &\boldmath{\abun}&  \boldmath{\abun} \\
                \hline 
                Ne\ione    & L & 8.03(19) & $-$0.12 &  4 &          & 8.15\\
                            & N & 7.81(20) &  $-$0.34 &  4 &              &     \\
                            
                Mg\ione    & L & 7.16(12) & $-$0.36 &  2 &          & 7.52\\
                           & N & 7.09(11) & $-$0.43 &  2 &                &     \\
                Mg\ii      & L & 6.82(26) & $-$0.70 &  9 & 7.07(08)&  \\ 
                           & N & 6.78(34) & $-$0.74 &  9 &          &   \\
                Al\ii      & L & 5.30(20) & $-$1.12 &  3 & 5.32(20) & 6.42 \\
                Si\ione    & L & 7.78(20) &  0.27 &  1 &            & 7.51\\
                           & N & 8.19(20) &  0.68 &  1 &               & \\
                Si\ii      & L & 8.72(15) &  1.21 & 20 & 8.42(22)& \\ 
                           & N & 8.75(15) &  1.24 & 20 &          &  \\ 
                Si\iii     & L & 9.45(13) &  1.94 &  3 & 9.49(13) & \\ 
                           & N & 9.26(09) &  1.75 &  3 &          &  \\
                P\ii       & L & 5.77(06) &  0.36 &  4 &          & 5.41\\
                S\ii       & L & 6.67(22) & $-$0.48 &  7 &          & 7.15 \\
                Cl\ii      & L & 7.10(11) &  1.85 &  2 &          & 5.25 \\
                Ca\ione    & L & 7.11(20) &  0.84 &  1 &      & 6.27\\
                           & N & 7.19(20) &  0.92 &  1 &          &  \\
                Ca\ii      & L & 6.81(33) &  0.54 &  5 &         &  \\ 
                           & N & 6.84(20) &  0.57 &  5 &           & \\ 
                Sc\ii      & L & 2.59(20) & $-$0.45 &  1 &          & 3.04\\
                           & N & 3.28(20) &  0.24 &  1 &          &  \\
                Ti\ii      & L & 5.32(22) &  0.42 & 19 & 5.91(14) & 4.90\\
                           & N & 5.33(23) &  0.43 & 19 &          &  \\ 
                V\ii       & L & 4.63(16) &  0.68 &  6 &           & 3.95\\
                Cr\ione    & L & 7.02(20) &  1.39 &  1 &        & 5.63\\
                Cr\ii      & L & 6.60(15) &  0.97 & 36 & 6.54(16) & \\ 
                Mn\ii      & L & 5.61(20) &  0.14 &  1 &          & 5.47 \\ 
                Fe\ione    & L & 7.72(10) &  0.27 &  7 & 7.54(25) & 7.45\\
                           & N & 7.80(06) &  0.35 &  7 &            &  \\ 
                Fe\ii      & L & 8.23(21) &  0.78 & 74 & 8.19(12) &  \\ 
                           & N & 8.26(20) &  0.81 & 74 &          & \\ 
                Fe\iii     & L & 9.33(28) &  1.88 & 12 & 9.06(30) & \\ 
                           & N & 9.32(30) &  1.87 & 12 &          &  \\
                Ni\ione    & L & 6.36(20) &  0.16 &  1 &        & 6.20\\
                Ni\ii      & L & 6.36(27) &  0.16 &  3 &          &  \\
                Sr\ii      & L & 5.23(04) &  2.35 &  2 &          & 2.88\\
                           & N & 5.06(03) &  2.18 &  2 &          &  \\ 
                Zr\ii      & L & 3.46(20) &  0.91 &  1 &          & 2.55\\
                           & N & 3.74(20) &  1.19 &  1 &          &   \\
                Ce\ii      & L & 4.72(27) &  3.14 &  5 &          & 1.58 \\
                Ce\iii     & L & 4.83(01) &  3.25 &  2 &          &  \\
                Pr\iii     & L & 4.11(31) &  3.39 & 28 &          & 0.72 \\
                Nd\ii      & L & 4.87(13) &  3.42 &  7 &          & 1.45 \\
                Nd\iii     & L & 4.81(32) &  3.36 & 60 &         & \\
                Sm\ii      & L & 4.81(20) &  3.86 &  1 &         & 0.95 \\
                Eu\ii      & L & 5.11(23) &  4.59 &  3 &          & 0.52 \\
                Eu\iii     & L & 4.92(21) &  4.40 &  3 &          &  \\
                Gd\ii      & L & 4.58(11) &  3.50 &  5 &          & 1.08 \\
                Tb\iii     & L & 3.57(21) &  3.26 &  5 &          & 0.31 \\
                Dy\iii     & L & 4.52(31) &  3.42 & 18 &         & \\
                Er\ii      & L & 4.08(11) &  3.15 &  2 &          &  0.93 \\
                Er\iii     & L & 3.60(33) &  2.67 &  7 &          &  \\
\noalign{\hrule height 1.0pt}
        \end{tabularx}
        \noindent{\footnotesize{L and N symbols indicate LTE and non-LTE mean abundances. n$_{\rm l}$  is the number of spectral lines used for abundance determination. The standard deviation is given in parentheses and was assumed to be 0.2 dex when the abundance was obtained from a single line. The abundances of MX TrA are converted to the form of \abun. The last column contains present-day solar system abundances taken from \cite{2021SSRv..217...44L}.}}

        \label{MeanAbunds}
\end{table}

\begin{figure}[H]
\includegraphics[width=12 cm]{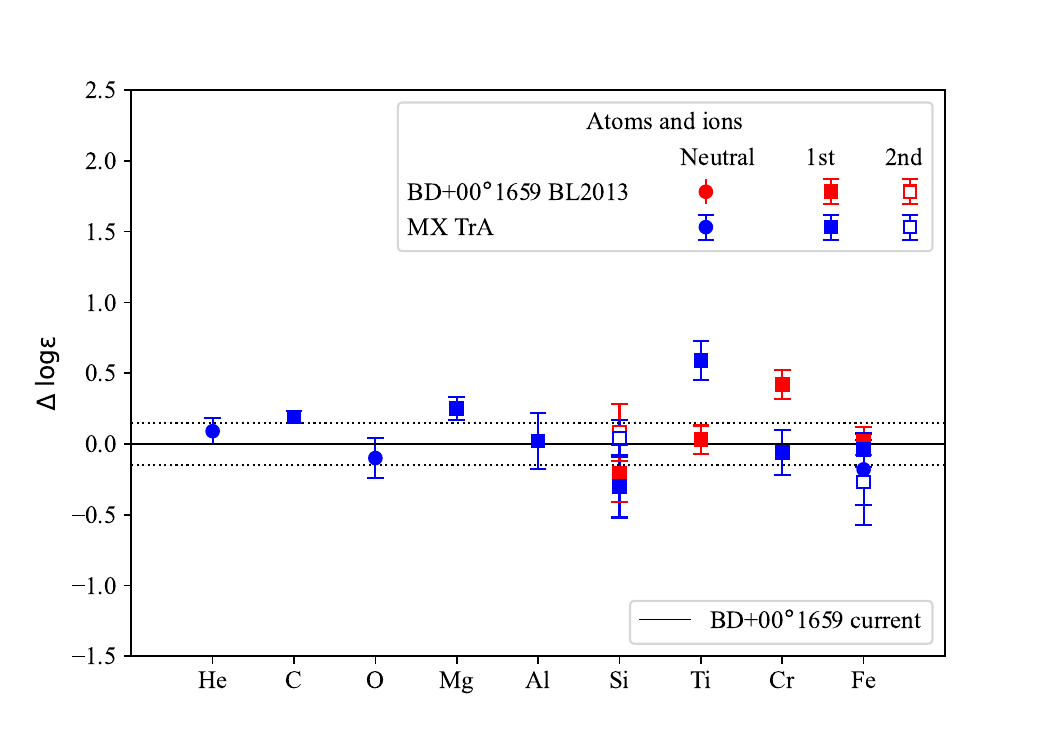}
\caption{Abundances in stars BD+00$^\circ$1659 \cite{2013A&A...551A..30B} and MX TrA  relative to the results of the present paper. Dotted lines indicates an average $\pm$0.15 dex abundance uncertainty.}
\label{fig4}
\end{figure}

Figure~\ref{fig4} shows a comparison of the derived abundances in BD+00$^\circ$1659 with those from \cite{2013A&A...551A..30B} and with the abundances of its twin star MX TrA (t11950g3.60) from \cite{2024MNRAS.52710376P}. Comparisons are made using LTE results except for O\ione\ where non-LTE values are available. Dotted lines indicate an average $\pm$0.15 dex uncertainty in abundance determination. Our LTE abundances well agree with those from \cite{2013A&A...551A..30B}. An exception is Cr, where we used the updated oscillator strength values from \cite{2017ApJS..228...10L}. Our abundances are consistent with those of MX TrA within the estimated uncertainty.

Although the predicted non-LTE corrections for Si and Fe reduce the abundance difference between lines of different ionisation stages, Si\ione--\iii\ and Fe\ione--\iii, this difference cannot be entirely removed. Therefore, we performed a stratification analysis for \mbox{both elements.}

\section{Stratification Analysis}\label{strat}
\subsection{Stratification Analysis of BD+00$^\circ$1659}
In magnetic Ap/Bp stars, the violation of ionisation balance serves as evidence for abundance stratification \cite{2003IAUS..210..301R}. Theoretical stratification models suggest that, as a first approximation, the vertical abundance distribution (stratification profile) can be represented by a step function \cite{1992A&A...263..232B}. This function is characterised by four parameters: the element abundances in the upper and lower atmospheric layers, the position of the abundance jump, and the width of this jump. These parameters are optimised to achieve the best fit with the observed profiles of spectral lines formed at different atmospheric layers.

The stratification analysis was performed with the IDL-based automated procedure \textsc{DDaFit} in conjunction with the \textsc{Synth3} code \cite{2007pms..conf..109K}. A precise stratification analysis requires selecting unblended or minimally blended spectral lines of atoms/first/second ions with different intensities and excitation potentials $E_i$. It is also crucial to choose spectral lines formed at different optical depths, particularly for Si, as the wings of strong Si\ii\ lines form at similar depths to Si\iii\ lines. Bailey and Landstreet \cite{2013A&A...551A..30B} noted that a Si abundance jump of approximately 1 dex at \lgt $\sim-0.4$ can reduce the observed Si\ii/Si\iii\ abundance difference to 0.3 dex. However, while a larger jump may remove the ionisation discrepancy, it does not allow fitting the wings of strong Si\ii\ lines. This reason guided our choice of Si lines for the stratification study, where we used one strong line of Si\ii\ at 4130.89~\AA\ and three lines of Si\iii. The Fe stratification analysis relies on four lines of Fe\ione, ten lines of Fe\ii, and five lines of Fe\iii. 

\begin{figure}[H]
     \includegraphics[width=0.7\textwidth]{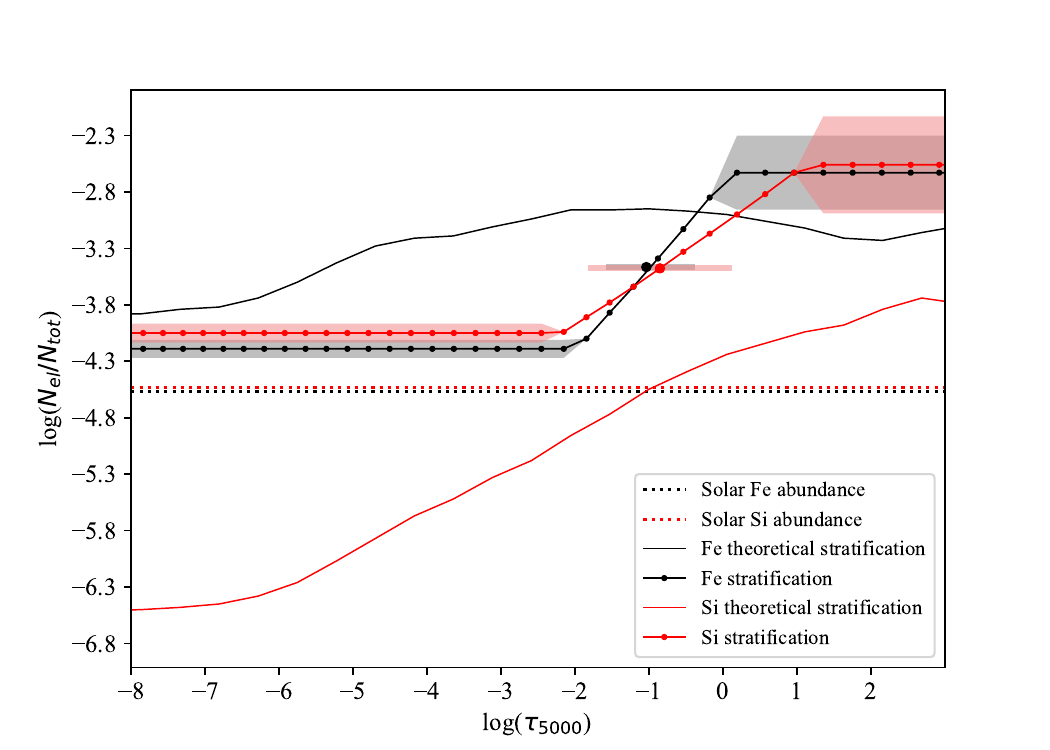}
  \caption{Distribution of Fe (black line) and Si (red line) abundance in the atmosphere of BD+00$^\circ$1659. Shaded areas show error bars for stratification parameters (higher and lower abundances and jump position). Theoretical predictions for the model \teff\ = 12,000 K, \lgg\ = 4.0 are shown by solid lines~\cite{2009A&A...495..937L}. Solar abundances are indicated by dotted lines.}
 \label{stratif-profiles}
 \end{figure}

\begin{figure}[H]

\begin{adjustwidth}{-\extralength}{0cm}
\centering 
 \hspace{-60pt}\begin{subfigure}{0.31\textwidth}
\centering
     \includegraphics[width=1.2\textwidth]{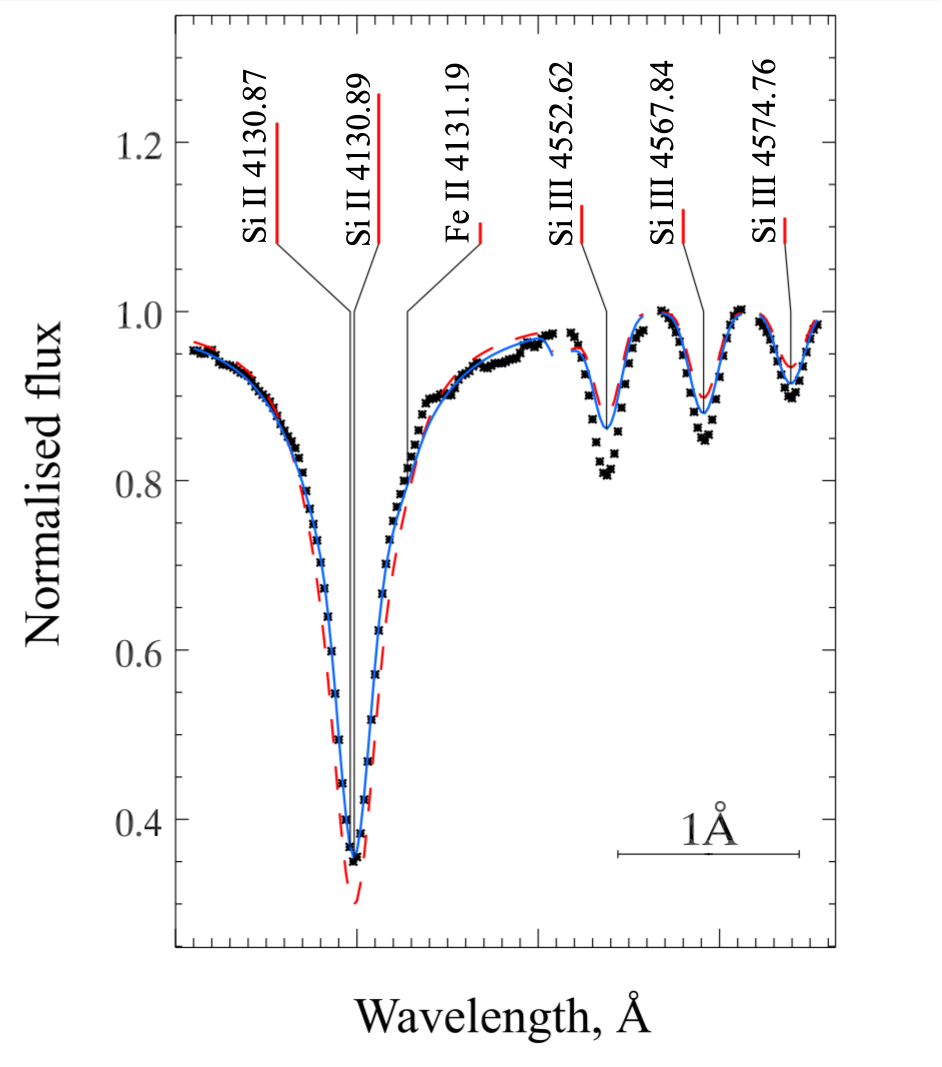}
     \caption{Lines of Si.}
     \label{si-lines}
 \end{subfigure}
 \quad\quad\quad
 \begin{subfigure}{0.67\textwidth}
\centering
     \includegraphics[width=1.2\textwidth]{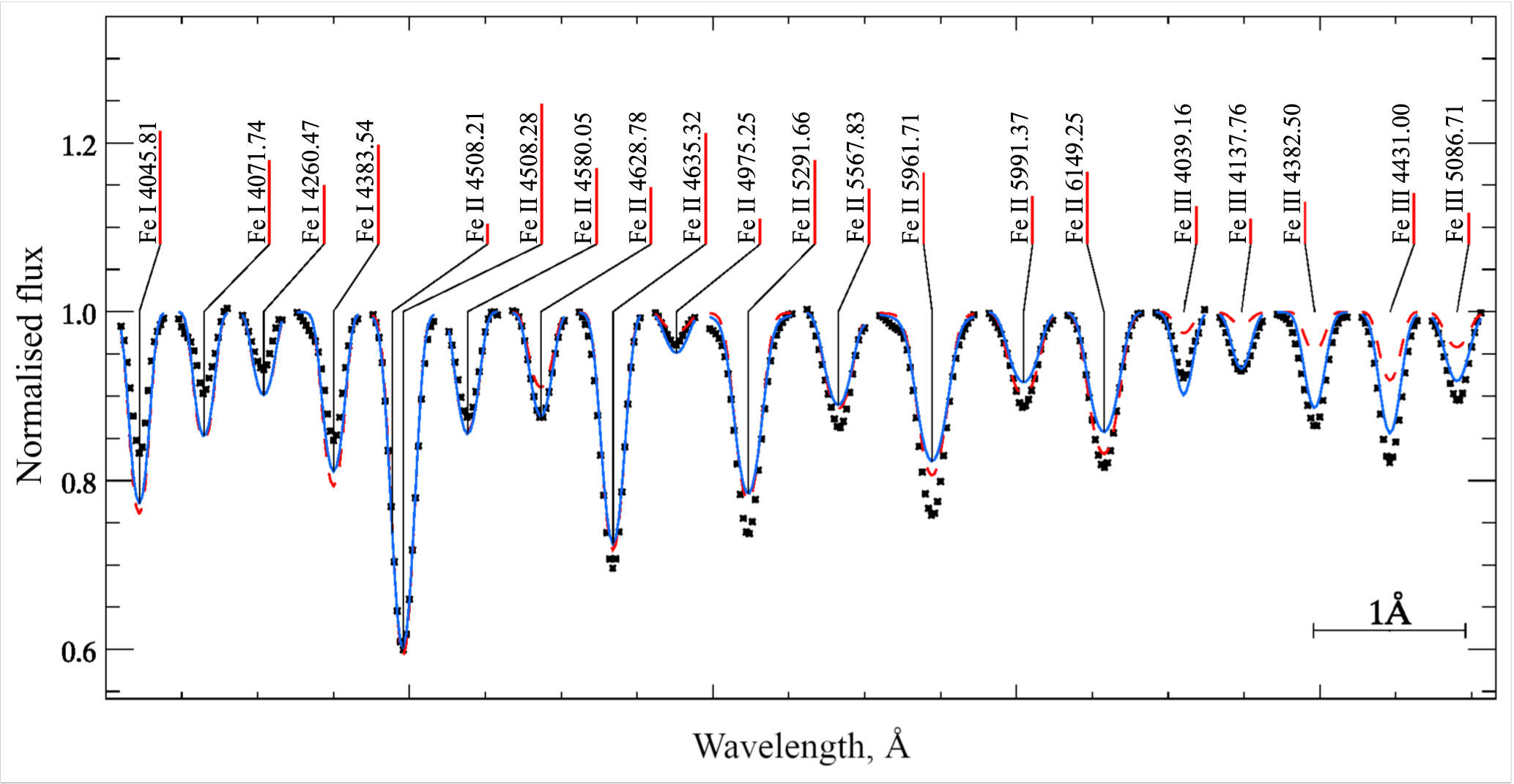}
     \caption{Lines of Fe.}
     \label{fe-lines}
 \end{subfigure}
\end{adjustwidth}
  \caption{Spectral line profiles used in stratification analysis. Black crosses indicate the observed profile. Synthetic spectra calculated with homogeneous abundance and with a stratified one are shown by red dashed and blue solid lines, respectively.}
  \label{fig6}
 \end{figure}

Figure \ref{stratif-profiles} displays the calculated stratification profiles of Si and Fe in the stellar atmosphere. Figure \ref{fig6}a, b show the fitted line profiles calculated with a homogeneous element distribution and when accounting for stratification. The impact of stratification on the Fe \ione\ and Fe\ii\ lines is minor, while stratification significantly improves the fitted Fe\iii\  line profile. For Si, stratification provides a better fit to the observed line profiles compared to those calculated with a depth-independent abundance distribution.

 Abundance jumps for both elements occur in atmospheric layers with an optical depth close to \lgt\ = $-$1.0 (see Table~\ref{strat-param}). The derived Si stratification profile aligns with the theoretical Si diffusion calculations of \cite{2009A&A...495..937L}, although the simple theory does not predict the large observed Si overabundance. For iron, the theoretical abundances in the upper and lower atmosphere are closer to the observed values, but the position of the theoretical jump is shifted to much higher atmospheric layers.
 
\vspace{-6pt}
 \begin{table}[H]
\caption{Stratification parameters of Fe and Si in the atmosphere of BD+00$^\circ$1659.}\label{strat-param} 
\tablesize{\small}
               \begin{tabularx}{\textwidth}{CCCCC}
\toprule
	\textbf{Element} & \boldmath{$log\left(\frac{N}{N_{tot}}\right)_{up}$} & \boldmath{$log\left(\frac{N}{N_{tot}}\right)_{low}$} & \boldmath{$log\tau_{5000}$} & \boldmath{$\triangle log\tau_{5000}$}    \\ 
\midrule
	Fe & ~$-$4.19(0.08) & $-$2.63(0.33) & $-$0.98(0.09) & 0.60(0.59)   \\ 
	Si & ~$-$4.05(0.08) & $-$2.56(0.43) & $-$0.85(0.16) & 0.98(0.54)   \\
\bottomrule
    \end{tabularx}
   \noindent{\footnotesize{The parameters of the step function of element stratification are presented, including the abundance in the upper and lower atmospheres, the position of the jump, and the width of the jump.}}
\end{table}

\begin{figure}[H]
\includegraphics[width=9 cm]{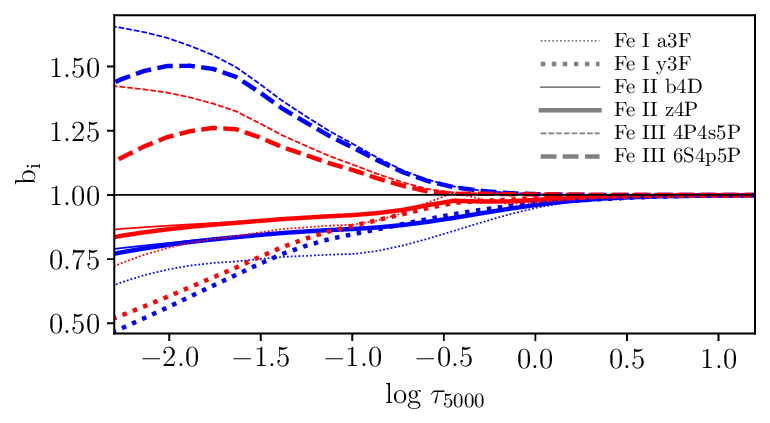}
\caption{The departure coefficients for the selected atomic levels of Fe\ione\, (dotted lines), Fe\ii\, (solid lines), and Fe\iii\, (dashed lines) as a function of optical depth in the model atmosphere of BD+00$^\circ$1659. See the legend for the levels' designation. Red and blue lines correspond to calculations accounting for and neglecting iron abundance stratification, respectively.}
\label{bi_fe}
\end{figure}

We estimated the impact of non-LTE effects on the abundance stratification for iron. Figure~\ref{bi_fe} shows the departure coefficients for selected levels of Fe\ione~4071~\AA, Fe\ii~6149~\AA, and Fe\iii~4431~\AA\ in a stratified and homogeneous model atmosphere. We found non-LTE effects to be minor when accounting for stratification, and the departure coefficients show smaller deviations from unity compared to calculations that neglect stratification. As a result, the iron line profiles of Fe\ione--\iii~ ionisation stages, calculated with a stratified model atmosphere in LTE and non-LTE, practically coincide (see Figure~\ref{stratif-lines-lte-nlte}). For iron and silicon, accounting for stratification shifts the formation depths of their spectral lines towards deeper atmospheric layers, where non-LTE effects are minor. Thus, in this study, we perform the abundance analysis of Si and Fe lines in LTE.

\vspace{-6pt}
 \begin{figure}[H]
     \includegraphics[width=0.8\textwidth]{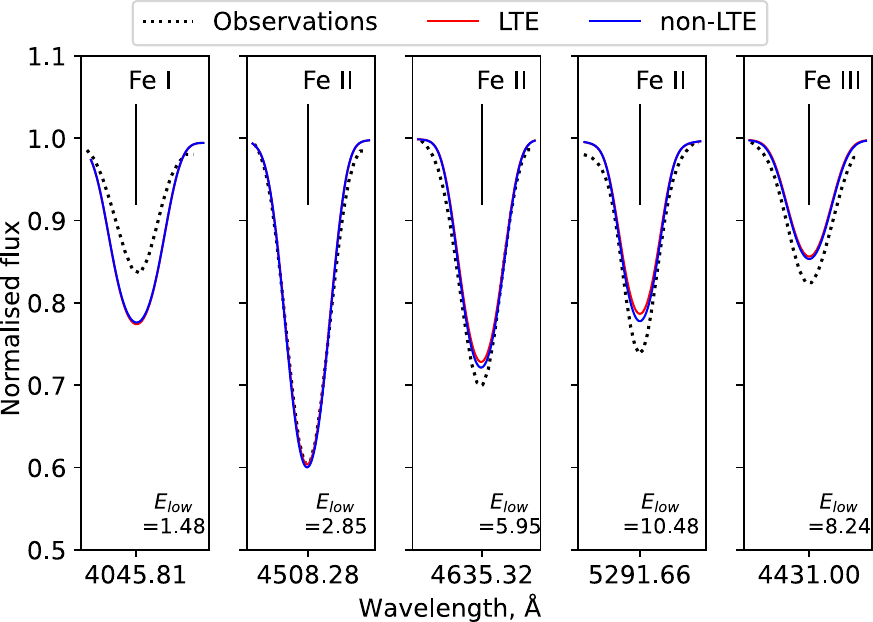}
  \caption{Comparison of Fe line profiles of different ionisation stages, calculated with stratified model in LTE and non-LTE. The excitation energy of the lower level ($E_{low}$) is indicated.}
 \label{stratif-lines-lte-nlte}
 \end{figure}

A horizontal magnetic field can impact elemental stratification in the upper atmosphere by reducing the effective diffusion coefficient of charged ions by a factor of two as the magnetic field strength increases \cite{2009A&A...495..937L}. However, since we assume a small (zero) magnetic field in BD+00$^\circ$1659, the possible impact of the magnetic field on stratification was not considered.

\subsection{Application of Stratification to MX TrA}\label{Comp}

To check the possibility of spreading stratification results to Ap/Bp stars with faster rotation, we applied the derived Fe and Si vertical abundance distributions to the ApSi star MX TrA, which is shown to be a twin of BD+00$^\circ$1659 (Sections~\ref{obs} and \ref{abund}) and also possesses strong silicon overabundance and a Si\ii--\iii\ ionisation imbalance in its atmosphere. Figure~\ref{strat_comp} illustrates the synthetic spectra calculated with the homogeneous and stratified model atmosphere t11950g3.60 fitted to the observed profiles of the selected Si\ii--\iii, Fe\ii--\iii\ lines in MX TrA. One can see that accounting for the BD+00$^\circ$1659 stratification profile results in a much better fit of the selected lines in the MX TrA spectrum than in the case of a chemically homogeneous atmosphere.

\begin{figure}[H]
\captionsetup[subfigure]{justification=centering}
 \begin{subfigure}[t]{1.0\textwidth}
\centering
     \includegraphics[width=\textwidth]{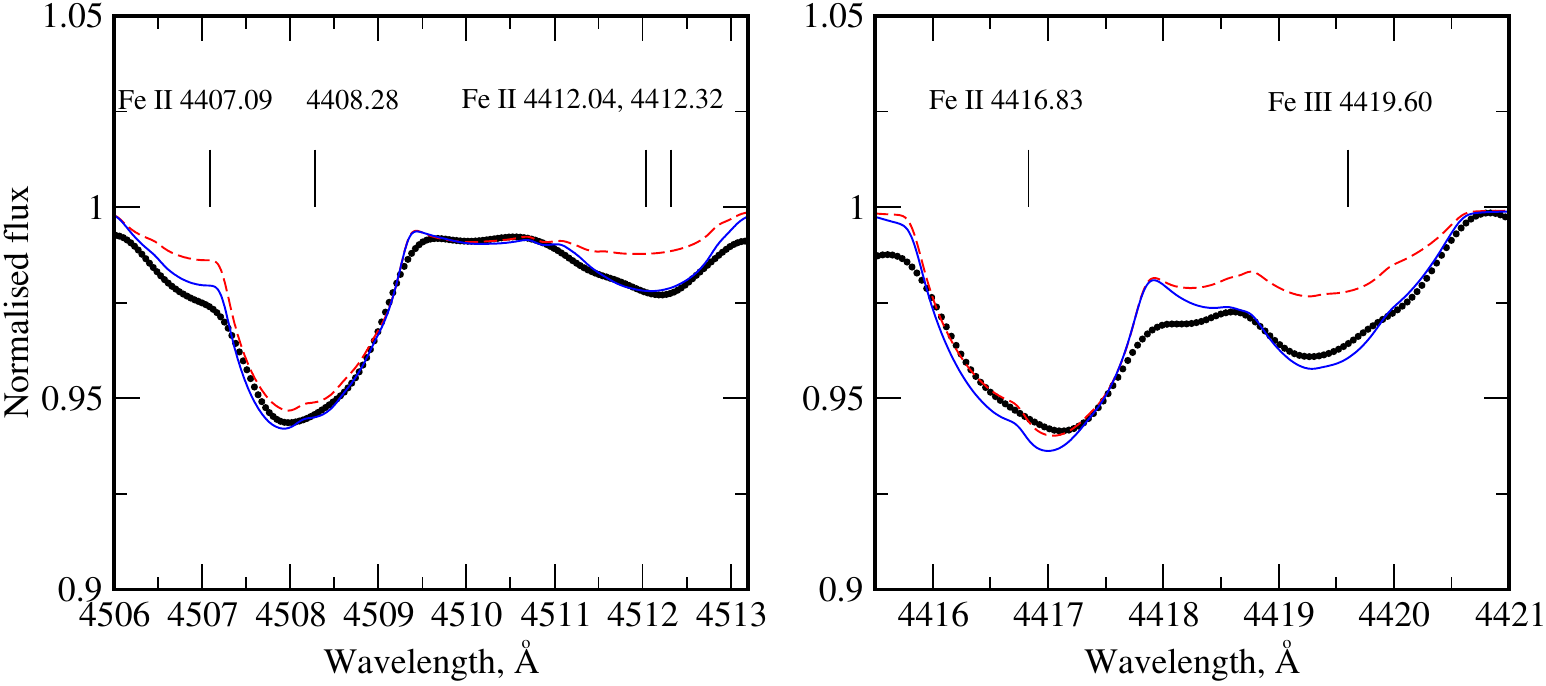}
     \caption{}
 \end{subfigure}

\caption{\textit{Cont.}}
\end{figure}
\begin{figure}[H]\ContinuedFloat

 \begin{subfigure}[t]{1.0\textwidth}
\centering
     \includegraphics[width=\textwidth]{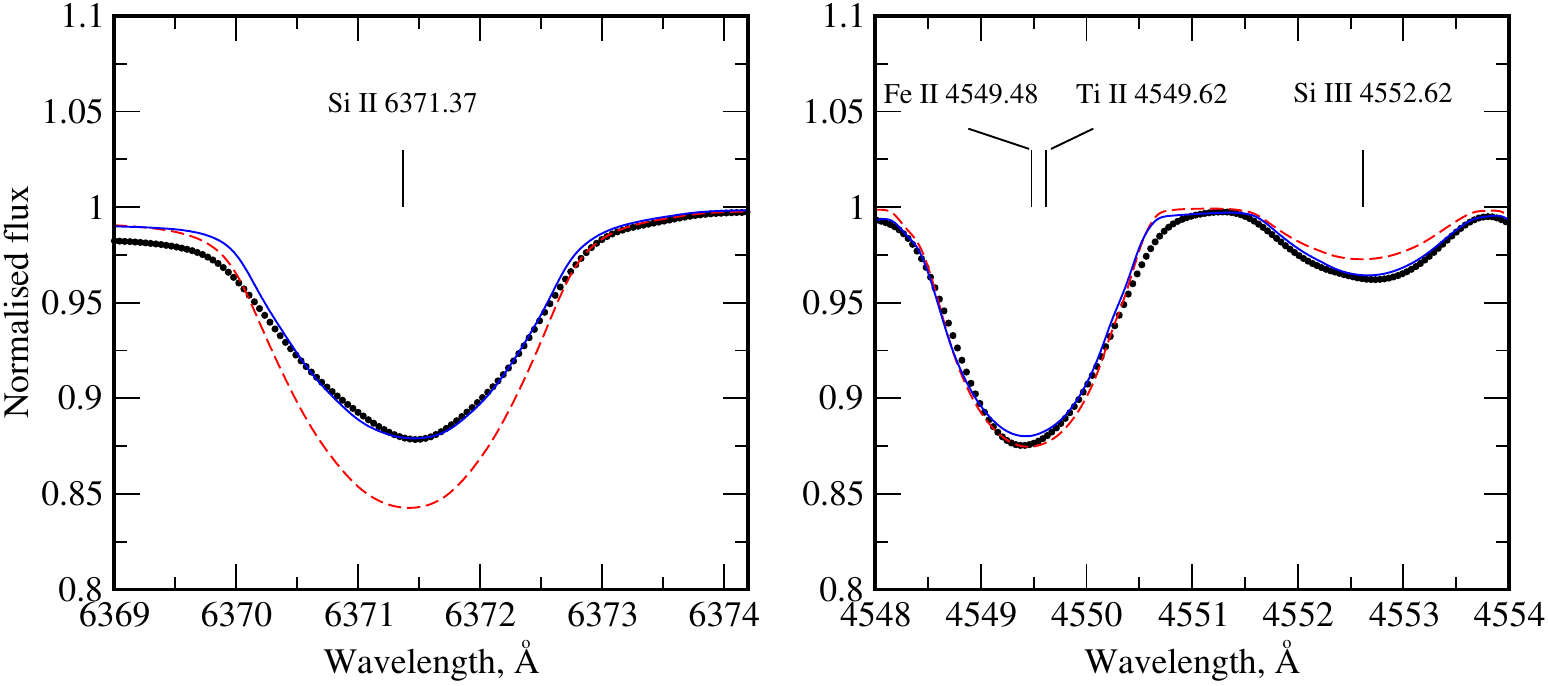}
     \caption{}
 \end{subfigure}
  \caption{Comparison between the observed Si (\textbf{a}) and Fe (\textbf{b}) line profiles in MX TrA (black filled circles) and synthetic spectra calculated with stratified (blue line) and homogeneous (red dashed line) model atmosphere.}\label{fig8}
 \label{strat_comp} 
\end{figure}

\section{Discussion}\label{disscussion}

\begin{figure}[H]
 \begin{subfigure}{0.49\textwidth}
\centering
     \includegraphics[width=\textwidth]{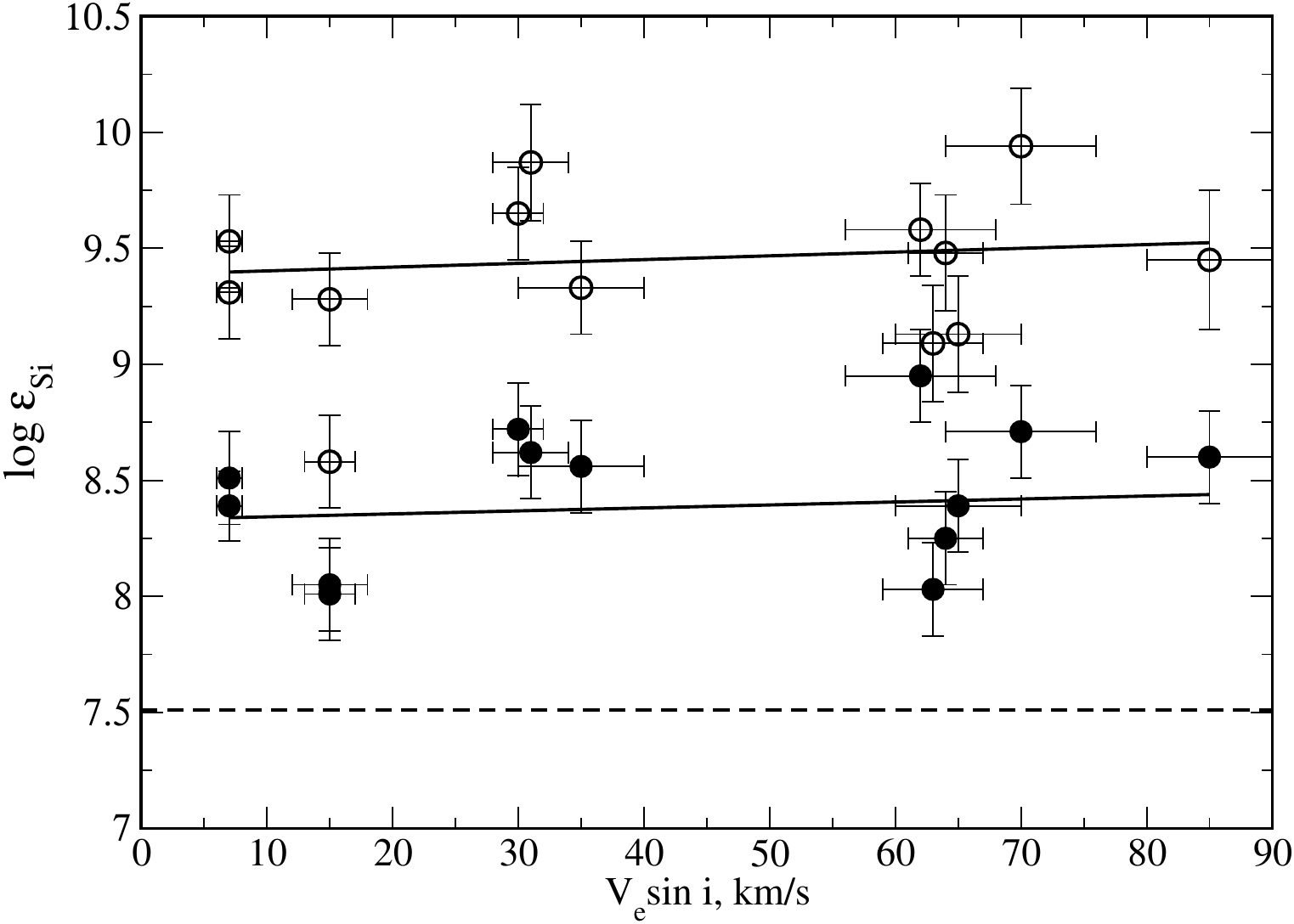}
 \end{subfigure}
 \hfill
 \begin{subfigure}{0.49\textwidth}
\centering
     \includegraphics[width=\textwidth]{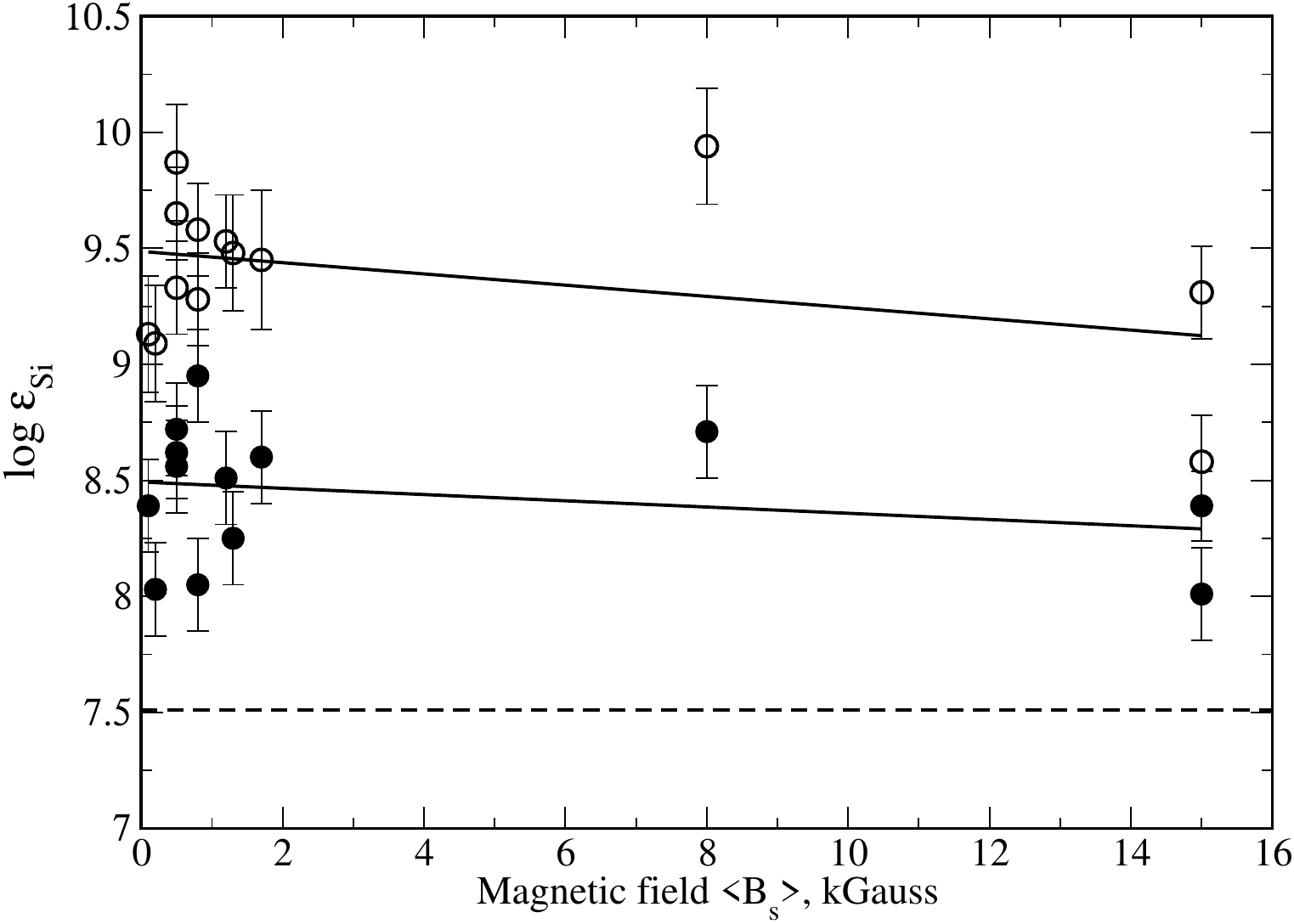}
 \end{subfigure}
  \caption{Si\ii\ (filled black circles) and Si\iii\ (open black circles) abundances from \cite{2013A&A...551A..30B} versus \vs\ and magnetic field. Solid lines show the linear regressions. The solar Si abundance is indicated by dashed line.}
   \label{vsini-magnfield}
 \end{figure}

Our analysis revealed that BD+00$^\circ$1659 is a representative member of the subgroup of chemically peculiar Ap/Bp stars with a chemically stratified atmosphere. We clearly showed that the spectra of two ApSi stars with close atmospheric parameters, slowly rotating BD+00$^\circ$1659 and faster rotating MX TrA, are perfectly represented by one model with abundance stratification. In theory, stellar rotation should disrupt atmospheric stability due to meridional circulation and inhibit element diffusion. However, in the spectrum of another ApSi star, 56 Ari (\vs\ = 160 \kms), Fe and Si overabundances of the same order are observed: \abun${Fe}$ = 7.9 and \abun${Si}$ = 8.5 \cite{2010A&A...509A..28S}. This indicates that element diffusion may produce similar abundance distributions in ApSi stars regardless of their rotational velocity. We analysed a sample of 17 ApSi stars with rotational velocities ranging from \mbox{7 to 85~\kms\ \cite{2013A&A...551A..30B}} and found no clear correlation between Si overabundance or Si\ii--\iii\ anomalies and \vs\ or between abundances and magnetic field strength. This is illustrated by Figure~\ref{vsini-magnfield}. Hence, we suppose that stratification is independent from the projected stellar rotational velocity (\vs) (within the limited temperature range) and consider BD+00$^\circ$1659 as a possible prototype for stratification studies of silicon \mbox{magnetic stars.}

Regarding the influence of stratification on non-LTE calculations, it is important to note that a comparable investigation was previously conducted only for neodymium (Nd) and praseodymium (Pr) \cite{2005A&A...441..309M, 2009A&A...495..297M} in cool Ap stars. In the atmospheres of these stars, Pr and Nd are concentrated in a thin upper atmospheric layer above \lgt~= $-$3.5, where non-LTE effects are rather large. The present stratification study demonstrates that Si and Fe settle in the deep atmospheric layers, where non-LTE effects are negligible. It seems reasonable to suggest that as the depth of line formation moves into deeper layers, the influence of non-LTE effects will decrease. Consequently, for such elements (Fe, Si, perhaps Mg, Ca, Sr), the LTE approach can be employed in stratification analysis.

\section{Conclusions}\label{conclusions}
We performed a self-consistent spectroscopic analysis of the slowly rotating ApSi star BD+00$^\circ$1659, deriving the abundances of 30 elements from He to Er and accounting for non-LTE effects for 11 of them. The star BD+00$^\circ$1659 displays the abundance pattern typical for the magnetic Ap/Bp Si stars.

The analysis of silicon and iron abundances using spectral lines from different ionisation states reveals a clear violation of ionisation balance, indicating the presence of abundance stratification in the stellar atmosphere. Stratification analysis of Si and Fe shows that these elements are concentrated in deeper atmospheric layers, below \lgt~= $-$1.0. We demonstrate that when elements are concentrated in these deeper layers, non-LTE effects are negligible, allowing the LTE approach to be successfully applied in \mbox{stratification analysis}. 

We showed that BD+00$^\circ$1659 has the same radius, atmospheric parameters, and elemental abundances as the more rapidly rotating ApSi star MX TrA. We found no clear correlation between Si overabundance or Si\ii--\iii\ anomalies and \vs\  in a sample of Ap/Bp Si stars from the literature. As a result, BD+00$^\circ$1659 can serve as a benchmark for studying the impact of stratification on emergent flux and lines profiles in MX TrA and other fast-rotating Ap/Bp Si stars within the  \teff\, = 11,000--13,000 temperature range, where severe line blending caused by rotation prevents direct stratification analysis.



\vspace{6pt} 




\authorcontributions{Conceptualisation, I.P. and T.R.; methodology, T.R.; software, Y.P. and T.S.; validation, A.R., T.R., and Y.P.; formal analysis, A.R. and T.R.; investigation, A.R., T.R., and T.S.; resources, A.R. and Y.P.; data curation, A.R.; writing---original draft preparation, A.R.; writing---review and editing, A.R., T.R., I.P., and T.S.; visualisation, A.R. and T.R.; supervision, I.P.; project administration, I.P. and T.R.; funding acquisition, I.P. All authors have read and agreed to the published version of the manuscript.}

\funding{This research was funded by the Russian Science Foundation (project No. 24-22-00237), https://rscf.ru/en/project/24-22-00237/ (accessed on 23 September 2024).}




\dataavailability {Spectra of BD+00$^\circ$1659 normalised to the continuum level are provided as supplementary material in the form of an ASCII file in CDS. The full table of used lines with $loggf$,  excitation potentials $E_i$, LTE and non-LTE abundances, and references for oscillator strengths and hyperfine constants is available online in CDS.} 




\acknowledgments{We acknowledge L. I. Mashonkina for providing the non-LTE calculations and valuable comments on this study.
}

\conflictsofinterest{The authors declare no conflicts of interest.} 



\abbreviations{Abbreviations}{
The following abbreviations are used in this manuscript:\\

\noindent 
\begin{tabular}{@{}ll}
REE & Rare-Earth Elements\\
ESPaDOnS & The Echelle SpectroPolarimetric Device for the Observation of Stars \\
PI & Principal Investigator\\
LTE & Local Thermodynamic Equilibrium

\end{tabular}
}




\begin{adjustwidth}{-\extralength}{0cm}

\reftitle{References}

\PublishersNote{}
\end{adjustwidth}
\end{document}